\documentclass[preprints,article,accept,moreauthors,pdftex,10pt,a4paper]{Definitions/mdpi}

\usepackage{dsfont}

\newcommand{\mR}{\mathbb{R}}

\newcommand{\mL}{\mathbb{L}}

\newcommand{\mZ}{\mathbb{Z}}
\newcommand{\cL}{\mathcal{L}}
\newcommand{\bq}{\begin{equation}}
\newcommand{\eq}{\end{equation}}
\newcommand{\mP}{\mathds{P}}

\newcommand{\cG}{\mathcal{G}}
\newcommand{\T}{\mathcal{T}}
\newcommand{\W}{\mathcal{W}}
\newcommand{\R}{\mathcal{R}}

\newcommand{\cP}{\mathcal{P}}
\newcommand{\I}{\mathcal{I}}

\firstpage{1} 
\pubvolume{xx}
\issuenum{1}
\articlenumber{5}
\pubyear{2018}
\copyrightyear{2018}
\history{Received: date; Accepted: date; Published: date}
\pdfoutput=1

\Title{Geometry of Thermodynamic Processes}

\Author{Arjan van der Schaft $^{1}$, and Bernhard Maschke $^{2}$}

\AuthorNames{Arjan van der Schaft, and Bernhard Maschke}

\address{%
$^{1}$ \quad Bernoulli Institute for Mathematics, Computer Science and Artificial Intelligence, Jan C. Willems Center for Systems and Control, University of Groningen, the Netherlands; a.j.van.der.schaft@rug.nl\\
$^{2}$ \quad Univ. Lyon 1, Universit\'e Claude Bernard Lyon 1, CNRS, LAGEP UMR 5007, Villeurbanne, France; bernhard.maschke@univ-lyon1.fr}

\corres{Correspondence: a.j.van.der.schaft@rug.nl; Tel.: +31-50-3633731}

\abstract{
Since the 1970s contact geometry has been recognized as an appropriate framework for the geometric formulation of the state properties of thermodynamic systems, without, however, addressing the formulation of non-equilibrium thermodynamic processes. In \cite{balian} it was shown how the symplectization of contact manifolds provides a new vantage point; enabling, among others, to switch between the energy and entropy representations of a thermodynamic system. In the present paper this is continued towards the global geometric definition of a degenerate Riemannian metric on the homogeneous Lagrangian submanifold describing the state properties, which is overarching the locally defined metrics of Weinhold and Ruppeiner. Next, a geometric formulation is given of non-equilibrium thermodynamic processes, in terms of Hamiltonian dynamics defined by Hamiltonian functions that are homogeneous of degree one in the co-extensive variables and zero on the homogeneous Lagrangian submanifold. The correspondence between objects in contact geometry and their homogeneous counterparts in symplectic geometry, as already largely present in the literature \cite{libermann,arnold}, appears to be elegant and effective. This culminates in the definition of {\it port-thermodynamic systems}, and the formulation of interconnection ports. The resulting geometric framework is illustrated on a number of simple examples, already indicating its potential for analysis and control.}


\begin{document}

\section{Introduction}
This paper is concerned with the {\it geometric formulation} of {\it thermodynamic systems}.
While the geometric formulation of {\it mechanical} systems has given rise to an extensive theory, commonly called {\it geometric mechanics},
the geometric formulation of thermodynamics has remained more elusive and restricted. 

Starting from Gibbs' fundamental relation, {\it contact geometry} has been recognized since the 1970s as an appropriate framework for the geometric formulation of thermodynamics; see in particular \cite{hermann, mrugala1, mrugala2, mrugala3, mrugala4}. More recently, the interest in contact-geometric descriptions has been growing, from different points of view and with different motivations; see e.g. \cite{eberard,merker,bravetti}, \cite{gromov, gromov1, yoshimura, favache09, favache10, ramirez1, ramirez2, ramirez3}. 
Despite this increasing interest, the current geometric theory of thermodynamics falls short in a number aspects. First, most of the work is on the geometric formulation of the {\it equations of state}, through the use of Legendre submanifolds \cite{hermann, mrugala1, mrugala2, mrugala3, mrugala4}, while not much attention has been paid to the geometric definition of non-equilibrium {\it dynamics}. Secondly, thermodynamic system models commonly appear both in {\it energy} and in {\it entropy} representation,
while in principle this corresponds to {\it different} contact-geometric descriptions. This is already demonstrated by rewriting Gibbs' equation in energy representation $dE=TdS -PdV$, with intensive variables $T, -P$, into the entropy representation $dS = \frac{1}{T} dE + \frac{P}{T} dV$, with intensive variables $\frac{1}{T}, \frac{P}{T}$. Thirdly, for reasons of analysis and control of composite thermodynamic systems, a geometric description of the {\it interconnection} of thermodynamic systems is desirable, but largely lacking. 

A new viewpoint on the geometric formulation of thermodynamic systems was provided in \cite{balian}, by exploiting the well-known result in geometry that odd-dimensional contact manifolds can be naturally {\it symplectized} to even-dimensional symplectic manifolds with an additional structure of {\it homogeneity}; see \cite{arnold, libermann} for textbook expositions. While the classical applications of symplectization are largely confined to time-dependent Hamiltonian mechanics \cite{libermann} and partial differential equations \cite{arnold}, the paper \cite{balian} argued convincingly that symplectization provides an insightful angle to the geometric modeling of thermodynamic systems as well. In particular, it yields a clear way to bring together energy and entropy representations, by viewing the choice of different intensive variables as the selection of different {\it homogeneous coordinates}. 

In the present paper we aim at expanding this symplectization point of view towards thermodynamics, amplifying on our initial work \cite{LHMNC18a, LHMNC18b}. In particular, we show how the symplectization point of view not only unifies the energy and entropy representation, but is also very helpful in describing the dynamics of thermodynamic processes, inspired by the notion of contact control system developed in \cite{eberard, favache09, favache10, ramirez1, ramirez2, ramirez3}, see also \cite{merker}. Furthermore, it yields a direct and global definition of a metric on the submanifold describing the state properties, encompassing the locally defined metrics of Weinhold \cite{weinhold} and Ruppeiner \cite{ruppeiner}. Finally, symplectization naturally leads to a definition of {\it interconnection ports}; thus extending the compositional geometric {\it port-Hamiltonian} theory (e.g. \cite{maschkevds, vanderschaftmaschkearchive, vds14}) of interconnected multi-physics systems to the thermodynamic realm.
All this will be illustrated by a number of simple, but instructive, examples, primarily serving to elucidate the developed framework and its potential.

\section{Thermodynamic phase space and geometric formulation of the equations of state}
Starting point for the geometric formulation of thermodynamic systems throughout this paper is an $(n+1)$-dimensional manifold $Q^e$, with $n\geq 2$, whose coordinates comprise the {\it extensive variables}, such as volume and mole numbers of chemical species as well as {\it entropy} and {\it energy} \cite{callen}. Emphasis in this paper will be on {\it simple} thermodynamic systems, with a single entropy and energy variable. Furthermore, for notational simplicity we will assume
\bq
Q^e= Q \times \mR \times \mR,
\eq
with $S \in \mR$ the entropy variable, $E\in \mR$ the energy variable, and $Q$ the $(n-1)$-dimensional manifold of remaining extensive variables (such as volume and mole numbers). 

In composite (i.e., compartmental) systems we may need to consider {\it multiple} entropies or energies; namely for each of the components. In this case $\mR \times \mR$ is replaced by $\mR^{m_S} \times \mR^{m_E}$, with $m_S$ denoting the number of entropies and $m_E$ the number of energies; see Example \ref{heatexch} for such a situation. This also naturally arises in the {\it interconnection} of thermodynamic systems, as will be discussed in Section \ref{sec:int}.

\medskip

Coordinates for $Q^e$ throughout will be denoted by $q^e=(q,S,E)$, with $q$ coordinates for $Q$ (the manifold of remaining extensive variables). Furthermore, we denote by $\T^*Q^e$ the cotangent bundle $T^*Q^e$ {\it without} its zero-section. Given local coordinates $(q,S,E)$ for $Q^e$, the corresponding natural cotangent bundle coordinates for $T^*Q^e$ and $\T^*Q^e$ are denoted by
\bq
(q^e,p^e)=(q,S,E,p,p_S,p_E),
\eq
where the co-tangent vector $p^e:=(p,p_S,p_E)$ will be called the vector of {\it co-extensive variables}. 

Following \cite{balian} the {\it thermodynamic phase space} $\mP (T^*Q^e)$ is defined as the {\it projectivization} of $\T^*Q^e$, i.e., as the fiber bundle over $Q^e$ with fiber at any point $q^e \in Q^e$ given by the projective space\footnote{Recall that elements of $\mP(T^*_{q^e} Q^e)$ are identified with rays in $T^*_{q^e}Q^e$, i.e., non-zero multiples of a non-zero cotangent vector.} $\mP(T^*_{q^e} Q^e)$. The corresponding projection will be denoted by $\pi: \T^*Q^e \to \mP (T^*Q^e)$.

It is well-known \cite{arnold, libermann} that $\mP(T^*Q^e)$ is a {\it contact manifold} of dimension $2n+1$. Indeed, recall \cite{arnold, libermann} that a {\it contact manifold} is an $(2n+1)$-dimensional manifold $N$ equipped with a maximally non-integrable field of hyperplanes $\xi$. This means that $\xi = \ker \theta \subset TN$ for a, possibly only locally defined, $1$-form $\theta$ on $N$ satisfying $\theta \wedge (d \theta)^n \neq 0$.
By Darboux's theorem \cite{arnold, libermann} there exist local coordinates (called {\it Darboux coordinates})
$q_0,q_1,\cdots,q_n,\gamma_1,\ \cdots, \gamma_n$
for $N$ such that locally
\bq
\label{2}
\theta= dq_0 - \sum_{i=1}^n \gamma_i dq_i
\eq
Then, in order to show that $\mP(T^*M)$ for any $(n+1)$-dimensional manifold $M$ is a contact manifold, consider the {\it Liouville one-form} $\alpha$ on the cotangent bundle $T^*M$, expressed in natural cotangent bundle coordinates for $T^*M$ as $\alpha = \sum_{i=0}^n p_idq_i$. Consider a neighborhood where $p_0\neq 0$, and define the {\it homogeneous} coordinates
\bq
\gamma_i = -\frac{p_i}{p_0}, \quad i=1, \cdots,n,
\eq
which, together with $q_0,q_1, \cdots,q_n$, serve as local coordinates for $\mP(T^*M)$. This results in the locally defined contact form $\theta$ as in \eqref{2} (with $\alpha = p_0 \theta$). The same holds on any neighborhood where one of the other coordinates $p_1,\cdots,p_n$ is different from zero, in which case division by the non-zero $p_i$ results in other homogeneous coordinates. This shows that $\mP(T^*M)$ is indeed a contact manifold. Furthermore \cite{libermann, arnold}, $\mP(T^*M)$ is the canonical contact manifold in the sense that every contact manifold $N$ is locally contactomorphic to $\mP(T^*M)$ for some manifold $M$. 

Taking $M=Q^e$, it follows that coordinates for the thermodynamical phase space $\mP(T^*Q^e)$ are obtained by replacing the coordinates $p^e=(p,p_S,p_E)$ for the fiber $T^*_{q^e} Q^e$ by {\it homogeneous} coordinates for the projective space $\mP(T^*_{q^e} Q^e)$. In particular, assuming $p_E \neq 0$ we obtain the homogeneous coordinates
\bq
\gamma=:\frac{p}{-p_E}, \: \gamma_S:=\frac{p_S}{-p_E},
\eq
defining the intensive variables of the {\it energy representation}. 
Alternatively, assuming $p_S \neq 0$ we obtain the homogeneous coordinates\footnote{See \cite{balian} for a discussion of $p_S$, or $p_E$, as a {\it gauge} variable.}
\bq
\widetilde{\gamma}=:\frac{p}{-p_S}, \; \widetilde{\gamma}_E:=\frac{p_E}{-p_S},
\eq
defining the intensive variables of the {\it entropy representation}. 

\begin{Example}\label{GAS}
Consider a mono-phase, single constituent, gas in a closed compartment, with volume $q=V$, entropy $S$ and internal energy $E$, satisfying Gibbs' relation $dE=TdS -PdV$. In the {\it energy} representation the intensive variable $\gamma$ is given by the {\it pressure} $-P$, and $\gamma_S$ is the {\it temperature} $T$. In the {\it entropy} representation, the intensive variable $\widetilde{\gamma}$ is equal to $\frac{P}{T}$, while $\widetilde{\gamma}_E$ equals the reciprocal temperature $\frac{1}{T}$. 
\end{Example}

In order to provide the geometric formulation of the {\it equations of state} on the thermodynamic phase space 
$\mP(T^*Q^e)$, we need the following definitions. First, recall that a submanifold $\cL$ of $\T^*Q^e$ is called a {\it Lagrangian} submanifold \cite{arnold, libermann} if the symplectic form $\omega:=d \alpha$ is zero restricted to $\cL$ and the dimension of $\cL$ is equal to the dimension of $Q^e$ (the maximal dimension of a submanifold restricted to which $\omega$ can be zero).
\begin{Definition}\label{laghom}
A {\it homogeneous Lagrangian submanifold} $\cL \subset \T^*Q^e$ is a Lagrangian submanifold with the additional property that 
\bq
(q^e,p^e) \in \cL \Rightarrow (q^e,\lambda p^e) \in \cL, \qquad \mbox{for every } 0\neq \lambda \in \mR
\eq

\end{Definition}
In Appendix, Proposition \ref{app:lag}, homogeneous Lagrangian submanifolds are geometrically characterized as submanifolds $\cL \subset \T^*Q^e$ of dimension equal to $\dim Q^e$, on which not only the symplectic form $\omega=d\alpha$ but also the Liouville one-form $\alpha$ is zero.

Importantly, homogeneous Lagrangian submanifolds of $\T^*Q^e$ are in one-to-one correspondence with {\it Legendre} submanifolds of $\mP(T^*Q^e)$.
Recall that a submanifold $L$ of a $(2n+1)$-dimensional contact manifold $N$ is a Legendre submanifold \cite{arnold, libermann} if the locally defined contact form $\theta$ is zero restricted to $L$, and the dimension of $L$ is equal to $n$ (the maximal dimension of a submanifold restricted to which $\theta$ can be zero). 

\begin{Proposition}[\cite{libermann}, Proposition 10.16]
Consider the projection $\pi: \T^*Q^e \to \mP (T^*Q^e)$. Then $L \subset \mP (T^*Q^e)$ is a Legendre submanifold if and only if $\cL:=\pi^{-1}L \subset \T^*Q^e$ is a homogeneous Lagrangian submanifold. Conversely, any homogeneous Lagrangian submanifold $\cL$ is of the form $\pi^{-1}L$ for some Legendre submanifold $L$.
\end{Proposition}

In the contact geometry formulation of thermodynamic systems \cite{hermann, mrugala1, mrugala2, mrugala3} the equations of state are formalized as Legendre submanifolds. In view of the correspondence with homogeneous Lagrangian submanifolds we arrive at the following

\begin{Definition}
Consider $Q^e$ and the thermodynamical phase space $\mP (T^*Q^e)$. The state properties of the thermodynamic system are defined by a homogeneous Lagrangian submanifold $\cL \subset \T^*Q^e$, and its corresponding Legendre submanifold $L \subset \mP (T^*Q^e)$.
\end{Definition}

The correspondence between Legendre and homogeneous Lagrangian submanifolds also implies the following characterization of {\it generating functions} for any homogeneous Lagrangian submanifold $\cL \subset \T^*Q^e$. This is based on the fact \cite{libermann, arnold} that any Legendre submanifold $L \subset N$ in Darboux coordinates $q_0,q_1,\cdots,q_n, \gamma_1, \cdots, \gamma_n$ for $N$ can be locally represented as
\bq
\label{20}
%\begin{array}{rcl}
L = \{(q_0,q_1,\cdots,q_n, \gamma_1, \cdots, \gamma_n) \mid q_0=F - \gamma_J\frac{\partial F}{\partial \gamma_J}, 
%\\[2mm]
\, q_J = - \frac{\partial F}{\partial \gamma_J}, \, \gamma_I = \frac{\partial F}{\partial q_I}\}
%\end{array}
\eq
for some partitioning $I \cup J=\{1, \cdots, n \}$ and some function $F(q_I,\gamma_J)$ (called a {\it generating function} for $L$), while conversely any submanifold $L$ as given in \eqref{20}, for any partitioning $I \cup J=\{1, \cdots, n \}$ and function $F(q_I,\gamma_J)$, is a Legendre submanifold. 

Given such a generating function $F(q_I,\gamma_J)$ for the Legendre submanifold $L$, we now define, assuming $p_0 \neq 0$ and substituting $\gamma_J=-\frac{p_J}{p_0}$,
\bq
\label{23}
G(q_0, \cdots,q_n,p_0,\cdots,p_n):=-p_0F(q_I, -\frac{p_J}{p_0})
\eq
Then a direct computation shows that
\bq
- \frac{\partial G}{\partial p_0} = F(q_I, -\frac{p_J}{p_0}) + p_0 \frac{\partial F}{\partial \gamma_J}(q_I, -\frac{p_J}{p_0})\frac{p_J}{p_0^2} = F(q_I, \gamma_J) - \frac{\partial F}{\partial \gamma_J}\gamma_J ,
\eq
implying, in view of \eqref{20}, that
\bq
\label{23b}
%\begin{array}{rcl}
\pi^{-1}(L) = \{((q_0,\cdots,q_n,p_0,\cdots,p_n) \mid 
%\\[2mm]
q_0=- \frac{\partial G}{\partial p_0}, \, q_J = - \frac{\partial G}{\partial p_J}, \, p_I = \frac{\partial G}{\partial q^I}\}
%\end{array}
\eq
In its turn this implies that $G$ as defined in \eqref{23} is a generating function for the homogeneous Lagrangian submanifold $\cL=\pi^{-1}(L)$. If instead of $p_0$ another coordinate $p_i$ is different from zero, then by dividing by this $p_i \neq 0$ we obtain a similar generating function. This is summarized in the following proposition.
\begin{Proposition}
Any Legendre submanifold $L$ can be locally represented as in \eqref{20}, possibly after renumbering the index set $\{0,1,\cdots,n\}$, for some partitioning $I \cup J=\{1, \cdots, n \}$ and generating function $F(q_I,\gamma_J)$, and conversely for any such $F(q_I,\gamma_J)$ the submanifold $L$ defined by \eqref{20} is a Legendre submanifold. 

Any homogeneous Lagrangian submanifold $\cL$ can be locally represented as in \eqref{23b} with generating function $G$ of the form \eqref{23}, and conversely for any such $G$ the submanifold \eqref{23b} is a homogeneous Lagrangian submanifold.
\end{Proposition}
Note that the generating functions $G$ as in \eqref{23} are homogeneous of degree $1$ in the variables $(p_0,\cdots,p_n)$; see the Appendix for further information regarding homogeneity.

The simplest instance of a generating function for a Legendre submanifold $L$ and its homogeneous Lagrangian counterpart $\cL$ occurs when the generating $F$ as in \eqref{20} only depends on $q_1, \cdots,q_n$. In this case, the generating function $G$ is given by
\bq
\label{22}
G(q_0, \cdots,q_n,p_0,\cdots,p_n)=-p_0F(q_1, \cdots,q_n),
\eq
with the corresponding homogeneous Lagrangian submanifold $\cL=\pi^{-1}(L)$ locally given as
\bq
%\begin{array}{rcl}
\cL  =  \{ (q_0,\cdots,q_n, p_0, \cdots, p_n) \mid q_0=F(q_1, \cdots,q_n), 
%\\[2mm]
\, p_1= - p_0\frac{\partial F}{\partial q_1}, \cdots, p_n= - p_0\frac{\partial F}{\partial q_n} \}
%\end{array}
\eq
A particular feature of this case is the fact that exactly one of the extensive variables, in the above $q_0$, is expressed as a function of all the others, i.e., $q_1,\cdots,q_n$. At the same time, $p_0$ is unconstrained, while the other co-extensive variables $p_1, \cdots, p_n$ are determined by $p_0,q_1, \cdots,q_n$. For a general generating function $G$ as in \eqref{23}, this is not necessarily the case. For example, if $J=\{1, \cdots, n \}$, corresponding to a generating function $-p_0F(\gamma)$, then $q_0, \cdots, q_n$ are all expressed as a function of the unconstrained variables $p_0, \cdots,p_n$.

\begin{Remark}
In the present paper homogeneity in the co-extensive variables $(p,p_S,p_E)$ is crucially used. This is different from homogeneity with respect to the {\it extensive} variables $(q,q_S,q_E)$, as occurring e.g. in the Gibbs-Duhem relations \cite{callen}. Homogeneity with respect to $(q,q_S,q_E)$ means that whenever $(q,q_S,q_E,p,p_S,p_E) \in \cL$, then also $(\mu q, \mu q_S,\mu q_E,p,p_S,p_E) \in \cL$ for any $\mu \neq 0$. 
%In particular, homogeneity with respect to the extensive variables implies that the intensive variables are {\it dependent}.
\end{Remark}

The two most important representations of a homogeneous Lagrangian submanifold $\cL \subset \T^*Q^e$, and its Legendre counterpart $L \subset \mP (T^*Q)$, are the {\it energy representation} and the {\it entropy representation}. In the first case, $\cL$ is represented, as in \eqref{22}, by a generating function of the form
\bq
-p_EE(q,S)
\eq
yielding the representation
\bq
\label{enerep}
\cL=\{(q,S,E,p,p_S,p_E) \mid E= E(q,S), p=-p_E \frac{\partial E}{\partial q}(q,S),  p_S=-p_E \frac{\partial E}{\partial S}(q,S) \}
\eq
In the second case (the entropy representation), $\cL$ is represented by a generating function of the form
\bq
-p_SS(q,E)
\eq
yielding the representation
\bq
\label{entrep}
\cL=\{(q,S,E,p,p_S,p_E) \mid S= S(q,E), p=-p_S \frac{\partial S}{\partial q}(q,E),  p_E=-p_S \frac{\partial S}{\partial E}(q,E) \}
\eq
Note that in the energy representation the independent extensive variables are taken to be $q$ and the entropy $S$, while the energy variable $E$ is expressed as a function of them. On the other hand, in the entropy representation the independent extensive variables are $q$ and the energy $E$, with $S$ expressed as a function of them. Furthermore, in the energy representation, the co-extensive variable $p_E$ is 'free', while instead in the entropy representation the co-extensive variable $p_S$ is free. In principle also {\it other} representations could be chosen, although we will not pursue this. For instance in Example \ref{GAS} one could consider a generating function $-p_VV(S,E)$ where the extensive variable $V$ is expressed as function of the other two extensive variables $S,E$.

\medskip

As already discussed in \cite{hermann,mrugala1} an important advantage of describing the state properties by a Legendre submanifold $L$, instead of by writing out the equations of state, is in providing a {\it global} and {\it coordinate-free} pint of view, allowing for an easy transition between different thermodynamic potentials. Also, possible {\it singularities} of the equations of state are typically not reflected in $L$ (that is, $L$ is still a smooth submanifold). As seen before \cite{balian}, the description by a homogeneous Lagrangian submanifold $\cL$ has the additional advantage of yielding an intrinsic way for switching between the energy and the entropy representation.
\begin{Remark}
\label{remarkcor}
Although the terminology 'thermodynamic phase space' for $\mP(T^*Q^e)$ may suggest that all points in $\mP(T^*Q^e)$ are {\it feasible} for the thermodynamical system, this is actually {\it not} the case. The {\it state properties} of the thermodynamic system are specified by the Legendre submanifold $L \subset \mP(T^*Q^e)$, and thus the actual 'state space' of the thermodynamic system at hand is this submanifold $L$; {\it not} the whole of $\mP(T^*Q^e)$. 

A proper analogy with the Hamiltonian formulation of mechanical systems would be as follows. Consider the phase space $T^*Q$ of a mechanical system with configuration manifold $Q$. Then the Hamiltonian $H: T^*Q \to \mR$  defines a Lagrangian submanifold $\cL_H$ of $T^*\left(T^*Q\right)$ given by the graph of the gradient of $H$. The homogeneous Lagrangian submanifold $\cL$ is analogous to $\cL_H$, while the symplectized thermodynamic phase space $\T^*Q^e$ is analogous to $T^*\left(T^*Q\right)$.)
\end{Remark}

\section{The metric determined by the equations of state}

In a series of papers starting with \cite{weinhold} Weinhold investigated the Riemannian metric that is locally defined by the Hessian matrix of the energy expressed as a (convex) function of the entropy and the other extensive variables\footnote{The importance of this Hessian matrix, also called the stiffness matrix, was already recognized in \cite{callen,tisza}.}. Similarly, Ruppeiner \cite{ruppeiner}, starting from the theory of fluctuations, explored the locally defined Riemannian metric given by minus the Hessian of the entropy expressed as a (concave) function of the energy and the other extensive variables. Subsequently, Mruga{\l}a \cite{mrugala2} reformulated both metrics as living on the Legendre submanifold $L$ of the thermodynamic phase space, and showed that actually these two metrics are locally equivalent (by a local contact transformation); see also \cite{bravetti}.

In this section, crucially exploiting the symplectization point of view, we provide a completely geometric and global definition of a degenerate pseudo-Riemannian metric on the homogeneous Lagrangian submanifold $\cL$ defining the equations of state. Only requirement is the assumption of existence of a torsion-free connection on the space  $Q^e$ of extensive variables. Furthermore, in a coordinate system in which the connection is {\it trivial} (i.e., its Christoffel symbols are all zero), this metric will be shown to reduce to Ruppeiner's locally defined metric once we use homogeneous coordinates corresponding to the entropy representation, and to Weinhold's locally defined metric by using homogeneous coordinates corresponding to the energy representation. Hence, we show that the metrics of Weinhold and Ruppeiner are just two different local representations of this {\it same} globally defined degenerate pseudo-Riemannian metric, thereby reinterpreting and generalizing the result in \cite{mrugala2}. 
%Furthermore, the definition of this sub-Riemannian metric is slightly more general than the, and can be extended to non-trivial torsion-free connections on the total space of extensive variables.

Recall \cite{abraham} that an (affine) \emph{connection} $\nabla$ on an $(n+1)$-dimensional manifold $M$ is defined as an assignment
\bq
(X , Y)  \longmapsto  \nabla_X Y
\eq
for any two vector fields $X,Y$,
which is $\mR$-bilinear and satisfies $\nabla_{fX} Y =f \nabla_X Y$
and $\nabla_X(fY)=f \nabla_XY + X(f)Y$, for any function $f$ on $M$. This implies that $\nabla_X Y
(q)$ only depends on $X(q)$ and the value of $Y$ along a curve which
is tangent to $X$ at $q$. 
In local coordinates $q$ for $M$ the connection is determined by its \emph{Christoffel symbols} $\Gamma^a_{bc} (q), \, a,b,c=0,\cdots, n$, defined by
\bq
\nabla_{\frac{\partial}{\partial q_b}} \frac{\partial}{\partial q_c} =
\sum_{a=0}^n\Gamma^a_{bc} (q) \frac{\partial}{\partial q_a} 
\eq
The connection is called {\it torsion-free} if
\bq
\nabla_X Y - \nabla_Y X = [X,Y]
\eq
for any two vector fields $X,Y$, or equivalently if its Christoffel symbols satisfy the symmetry property $\Gamma^a_{bc} (q)= \Gamma^a_{cb} (q),\, a,b,c=0,\cdots, n$. We call a connection {\it trivial} in a given set of coordinates $q=(q_0,\cdots,q_n)$ if its Christoffel symbols in these coordinates are all zero.

As detailed in \cite{yano}, given a torsion-free connection on $M$ there exists a natural pseudo-Riemannian\footnote{'Pseudo' since the metric is indefinite.} metric on the cotangent-bundle $T^*M$, in cotangent bundle coordinates $(q,p)$ for $T^*M$ given as
\bq
2\sum_{i=0}^{n} dq_i \otimes dp_i - 2\sum_{a,b,c=0}^n p_c \Gamma^c_{ab} (q)dq_a \otimes dq_b
\eq
Let us now consider for $M$ the manifold of extensive variables $Q^e=Q \times \mR^2$ with coordinates $q^e=(q,S,E)$ as before, where we assume the existence of a torsion-free connection, which is trivial in the coordinates $(q,S,E)$, i.e., the Christoffel symbols are all zero. Then the pseudo-Riemannian metric $\I$ on $\T^*Q^e$ takes the form
\bq
\I:=2(dq \otimes dp + dS \otimes dp_S + dE \otimes dp_E) 
\eq
Denote by $\cG$ the pseudo-Riemannian metric $\I$ {\it restricted} to the homogeneous Lagrangian submanifold $\cL$ describing the state properties. Consider the {\it energy} representation \eqref{enerep} of $\cL$, with generating function $-p_E E(q,S)$. It follows that $\frac{1}{2}\cG$ equals (in shorthand notation)
\bq
\begin{array}{l}
dq \otimes d\left(-p_E \frac{\partial E}{\partial q}\right) + dS \otimes d\left(-p_E \frac{\partial E}{\partial S}\right) + dE \otimes dp_E = \\[2mm]
\; -p_Edq \otimes \left(\frac{\partial^2 E}{\partial q^2}dq +  \frac{\partial^2 E}{\partial q \partial S}dS \right) - dq \otimes \frac{\partial E}{\partial q} dp_E \\[2mm]
\; -p_EdS \otimes \left(\frac{\partial^2 E}{\partial q \partial S}dq +  \frac{\partial^2 E}{\partial S^2}dS \right) - dS \otimes \frac{\partial E}{\partial S} dp_E\\[2mm]
\; + \frac{\partial^T E}{\partial q} dq \otimes dp_E + \frac{\partial^T E}{\partial S} dS \otimes dp_E \\[2mm]
= -p_E \left(dq \otimes \frac{\partial^2 E}{\partial q^2}dq +  dq \otimes \frac{\partial^2 E}{\partial q \partial S}dS + dS \otimes \frac{\partial^2 E}{\partial q \partial S}dq +  dS \otimes \frac{\partial^2 E}{\partial S^2}dS \right) \\[2mm]
=: -p_E \W
\end{array}
\eq
where
\bq 
\label{weinhold}
\W = dq \otimes \frac{\partial^2 E}{\partial q^2}dq +  dq \otimes \frac{\partial^2 E}{\partial q \partial S}dS + dS \otimes \frac{\partial^2 E}{\partial S \partial q}dq +  dS \otimes \frac{\partial^2 E}{\partial S^2}dS
\eq
is recognized as Weinhold's metric \cite{weinhold}; the (positive-definite) Hessian of $E$ expressed as a (strongly convex) function of $q$ and $S$. 

On the other hand, in the {\it entropy} representation \eqref{entrep} of $\cL$, with generating function $-p_S S(q,E)$, an analogous computation shows that $\frac{1}{2}\cG$ is given as $p_S \R$, with
\bq
\label{ruppeiner}
\R = - dq \otimes \frac{\partial^2 S}{\partial q^2}dq -  dq \otimes \frac{\partial^2 S}{\partial q \partial E}dE - dE \otimes \frac{\partial^2 }{\partial E \partial q}dq -  dE \otimes \frac{\partial^2 S}{\partial E^2}dE
\eq
the Ruppeiner metric \cite{ruppeiner}; minus the Hessian of $S$ expressed as (strongly concave) function of $q$ and $E$. Hence we conclude that
\bq
-p_E \W = p_S \R,
\eq
implying $\W = -\frac{p_S}{p_E} \R= \frac{\partial E}{\partial S}\R =T \, \R$, with $T$ the temperature. This is basically the relation between $\W$ and $\R$ found in \cite{mrugala2}. Summarizing, we have found
\begin{Theorem}
Consider a torsion-free connection on $Q^e$, with coordinates $q^e=(q,S,E)$ in which the Christoffel symbols of the connection are all zero. Then by restricting the pseudo-Riemannian metric $\I$  to $\cL$ we obtain a degenerate pseudo-Riemannian metric $\cG$ on $\cL$, which in local energy-representation \eqref{enerep} for $\cL$ is given by $-2p_E \W$, with $\W$ the Weinhold metric \eqref{weinhold}, and in a local entropy representation \eqref{entrep} by $2p_S \R$, with $\R$ the Ruppeiner metric \eqref{ruppeiner}.
\end{Theorem}
We emphasize that the degenerate pseudo-Riemannian metric $\cG$ is {\it globally defined} on $\cL$, in contrast to the locally defined Weinhold and Ruppeiner metrics $\W$ and $\R$; see also the discussion in \cite{mrugala2, mrugala3,bravetti}. We refer to $\cG$ as {\it degenerate}, since its rank is at most $n$ instead of $n+1$. Note furthermore that $\cG$ is homogeneous of degree $1$ in $p^e$, and does {\it not} project to the Legendre submanifold $L$.

\medskip

This can be directly extended to non-trivial torsion-free connections $\nabla$ on $Q^e$. For example, consider the following situation.
For ease of notation denote $q_S:=S, q_E:=E$, and correspondingly denote $(q,S,E)= (q_0,q_1, \cdots,q_{n-2},q_S,q_E)$. Now, consider a torsion-free connection on $Q^e$ given by symmetric Christoffel symbols $\Gamma^c_{ab}=\Gamma^c_{ba}$, with indices $a,b,c=0, \cdots, n-2, S,E$, satisfying $\Gamma^c_{ab}=0$ whenever one of the indices $a,b,c$ is equal to the index $E$. Then the indefinite metric $\I$
on $\T^*Q^e$ is given by (again in shorthand notation)
\bq
2\sum_{i=0}^{E} dq_i \otimes dp_i - 2\sum_{a,b,c=0}^{S} p_c \Gamma^c_{ab} (q)dq_a \otimes dq_b
\eq
%\bq
%\begin{bmatrix} 0 & 1 & 0_{1,n+1} & 0_{1,n+1} \\
%1 & 0 & 0_{1,n+1} & 0_{1,n+1} \\
%0_{n+1,1} & 0_{n+1,1} & -2p_c\Gamma^c_{ab} & I_{n+1,n+1} \\
%0_{n+1,1} & 0_{n+1,1} & I_{n+1,n+1} & 0_{n+1,n+1} 
%\end{bmatrix}
%\eq
It follows that the resulting metric $\frac{1}{2}\cG$ on $\cL$ is given by the matrix 
%(with $\tilde{q}:= (q,S)$)
\bq
-p_E \left( \frac{\partial^2 E}{\partial q_a \partial q_b} - \sum_{c=0}^S \frac{\partial E}{\partial q_c}\Gamma^c_{ab}  \right)_{a,b=0,\cdots,S}
\eq
Here the $(n \times n)$-matrix at the right-hand side of $-p_E$ is the globally defined {\it geometric Hessian matrix} \cite{bullo} with respect to the connection on $Q \times \mR$ corresponding to the Christoffel symbols $\Gamma^c_{ab}$, $a,b,c=0, \cdots, n-2, S$.

\section{Dynamics of thermodynamic processes}

In this section we explore the geometric structure of the {\it dynamics} of (non-equilibrium) thermodynamic processes; in other words, geometric thermo{\it dynamics}. By making crucially use of the symplectization of the thermodynamic phase space this will lead to the definition of {\it port-thermodynamic systems} in Definition \ref{portthermo}; allowing for {\it open} thermodynamic processes. The definition is illustrated in Section \ref{subsec:ex} on a number of simple examples. In Section \ref{subsec:cont} initial observations will be made regarding the {\it controllability} of port-thermodynamic systems.

\subsection{Port-thermodynamic systems}
In Section 2 we noted the one-to-one correspondence between Legendre submanifolds $L $ of the thermodynamic phase space $\mP(T^*Q^e)$ and homogeneous Lagrangian submanifolds $\cL$ of the symplectized space $\T^*Q^e$. In the present section we start by noting that there is as well a one-to-one correspondence between {\it contact vector fields} on $\mP(T^*Q^e)$ and Hamiltonian vector fields $X_K$ on $\T^*Q^e$ with Hamiltonians that are {\it homogeneous} of degree $1$. (See the Appendix for further details on homogeneity.) 

Here, Hamiltonian vector fields $X_K$ on $\T^*Q^e$ with Hamiltonian $K$ are in cotangent bundle coordinates $(q^e,p^e)$ for $T^*Q^e$ given by the standard expressions
\bq
\dot{q}_i = \frac{\partial K}{\partial p_i}(q^e,p^e), \quad \dot{p}_i = -\frac{\partial K}{\partial q_i}(q^e,p^e), \quad i=0,1,\cdots,n ,
\eq
while {\it contact vector fields} $X_{\widehat{K}}$ on the contact manifold $\mP(T^*Q^e)$ are given in local Darboux coordinates $(q^e,\gamma)=$$(q_0,\cdots,q_n,\gamma_1,\cdots,\gamma_n)$ as
%\footnote{Here we follow the sign convention of \cite{arnold}, as opposed to the one in \cite{libermann}.} 
\cite{libermann,arnold}
\bq
\label{contactlocal}
\begin{array}{rcll}
\dot{q}_0 & = & \widehat{K}(q^e,\gamma) - \sum_{j=1}^n \gamma_j \frac{\partial \widehat{K}}{\partial \gamma_j}(q^e,\gamma) &\\[2mm]
\dot{q}_i &= &  - \frac{\partial \widehat{K}}{\partial \gamma_i}(q^e,\gamma), \quad & i=1, \cdots,n
\\[2mm]
\dot{\gamma}_i &= &  \frac{\partial \widehat{K}}{\partial q_i}(q^e,\gamma) + \gamma_i \frac{\partial \widehat{K}}{\partial q_0}(q^e,\gamma), \quad &
i=1, \cdots,n ,
\end{array}
\eq
for some {\it contact Hamiltonian} $\widehat{K}(q^e,\gamma)$.

Indeed, consider any Hamiltonian vector field $X_K$ on $\T^*Q^e$, with $K$ homogeneous of degree $1$ in the co-extensive variables $p^e$. Equivalently (see Appendix, Proposition \ref{prop1}), $\mL_{X_K} \alpha=0$. It follows, cf. Theorem 12.5 in \cite{libermann}, that $X_K$ projects under $\pi: \T^*Q^e \to \mP(T^*Q^e)$ to a vector field $\pi_*X_K$, satisfying
\bq
\mL_{\pi_*{X_K}} \theta = \rho \theta
\eq
for some function $\rho$, for all (locally defined) expressions of the contact form $\theta$ on $\mP(T^*Q^e)$. This exactly means \cite{libermann} that the vector field $\pi_*{X_K}$ is a contact vector field with contact Hamiltonian 
\bq
\label{Khat1}
\widehat{K}:=\theta (\pi_*{X_K})
\eq
Conversely \cite{libermann, arnold}, any contact vector field $X_{\widehat{K}}$ on $\mP(T^*Q^e)$, for some contact Hamiltonian $\widehat{K}$, can be {\it lifted} to a Hamiltonian vector field $X_K$ on $\T^*Q^e$ with homogeneous $K$. In fact, for $\widehat{K}$ expressed in Darboux coordinates for $\mP(T^*Q^e)$ as $\widehat{K}(q_0,q_1, \cdots, q_n,\gamma_1,\cdot,\gamma_n)$ the corresponding homogeneous function $K$ is given as, cf. \cite{libermann} (Chapter V, Remark 14.4),
\bq
\label{Khat}
K(q_0, \cdots, q_n,p_0,\cdots,p_n)= p_0 \widehat{K}(q_0, \cdots, q_n,-\frac{p_1}{p_0},\cdots, -\frac{p_n}{p_0}),
\eq
and analogously on any other homogeneous coordinate neighborhood of $\mP (T^*Q^e)$. This is summarized in the following proposition.\\
({\bf N.B.}: For brevity, we will {\it throughout} refer to a function $K(q^e,p^e)$ that is homogeneous of degree $1$ in the co-extensive variables $p^e$ as a {\it homogeneous function}, and to a Hamiltonian vector field $X_K$ on $\T^*Q^e$ with $K$ homogeneous of degree $1$ in $p^e$ as a {\it homogeneous Hamiltonian vector field}. )
\begin{Proposition}\label{homvf}
Any homogeneous Hamiltonian vector field $X_K$ on $\T^*Q^e$ projects under $\pi$ to a contact vector field $X_{\widehat{K}}$ on $\mP(T^*Q^e)$ with $\widehat{K}$ locally given by \eqref{Khat1}, and conversely any contact vector field $X_{\widehat{K}}$ on $\mP(T^*Q^e)$ lifts under $\pi$ to a homogeneous Hamiltonian vector field $X_K$ on $\T^*Q^e$ with $K$ locally given by \eqref{Khat}.
\end{Proposition}

Recall, see also Remark \ref{remarkcor}, that the equations of state describe the {\it constitutive} relations between the extensive and intensive variables of the thermodynamic system, or said otherwise, the {\it state properties} of the thermodynamic system. Since these properties are {\it fixed} for a given thermodynamic system any dynamics should leave its equations of state {\it invariant}. Equivalently, any dynamics on $\T^*Q^e$ or on $\mP (T^*Q^e)$ should leave the homogeneous Lagrangian submanifold $\cL \subset \T^*Q^e$, respectively, its Legendre submanifold counterpart $L \subset \mP (T^*Q^e)$, invariant\footnote{Recall that a submanifold is invariant for a vector field if the vector field is everywhere tangent to it; and thus solution trajectories remain on it.}. Furthermore, it is natural to require the dynamics of the thermodynamic system to be {\it Hamiltonian}; i.e., homogeneous Hamiltonian dynamics on $\T^*Q^e$ and a contact dynamics on $\mP (T^*Q^e)$. 

In order to combine the Hamiltonian structure of the dynamics with invariance we make crucially use of the following properties.
\begin{Proposition}\label{propinv}
\begin{enumerate}
\item
A homogeneous Lagrangian submanifold $\cL \subset \T^*Q^e$ is invariant for the homogeneous Hamiltonian vector field $X_K$ if and only if the homogeneous $K:\T^*Q^e \to \mR$ restricted to $\cL$ is zero. 
\item
A Legendre submanifold $L \subset \mP (T^*Q^e)$ is invariant for the contact vector field $X_{\widehat{K}}$ if and only if $\widehat{K}: \mP (T^*Q^e) \to \mR$ restricted to $L$ is zero.
\item
$K$ {\it restricted} to $\cL$ is zero if and only the function $\widehat{K}: \mP (T^*Q^e) \to \mR$ restricted to $L$ is zero. 

\end{enumerate}
\end{Proposition}
Item $2$ is well-known \cite{libermann,arnold}, item $1$ can be found in \cite{LHMNC18a,libermann}, while item $3$ directly follows from the correspondence between $K$ and $\widehat{K}$ in \eqref{Khat1} and \eqref{Khat}.

Based on these considerations we define the dynamics of a thermodynamic system as being produced by a homogeneous Hamiltonian function, parametrized by $u \in \mR^m$,
\bq
K:= K^a + K^cu : \T^*Q^e \to \mR, \quad u \in \mR^m,
\eq
with $K^a$ restricted to $\cL$ zero, and $K^c$ an $m$-dimensional row of functions $K^c_{j}, j=1, \cdots,m$, all of which are also zero on $\cL$. Then the resulting dynamics is given by the homogeneous Hamiltonian dynamics\footnote{In \cite{LHMNC18a, LHMNC18b} this was called a homogeneous Hamiltonian control system.} on $\T^*Q^e$
\bq
\dot{x} = X_{K^a}(x) + \sum_{j=1}^m X_{K^c_j}(x)u_j, \quad x=(q^e,p^e),
\eq
restricted to $\cL$. By Proposition \ref{homvf} this dynamics projects to contact dynamics corresponding to the contact Hamiltonian $\widehat{K} = \widehat{K}^a + \widehat{K}^cu$ on the corresponding Legendre submanifold $L \subset \mP(T^*Q^e)$.

The invariance conditions on the parametrized Hamiltonian $K$ defining the dynamics on $\cL$ and $L$ can be seen to take the following explicit form. Since $K$ is homogeneous of degree $1$ we can write by Euler's homogeneous function theorem (Theorem \ref{Euler})
\bq
\label{KaKc}
\begin{array}{rcll}
K^a &= &p^Tf +p_Sf_S + p_Ef_E, \quad & f=\frac{\partial K^a}{\partial p},  \, f_S=\frac{\partial K^a}{\partial p_S}, \, f_E=\frac{\partial K^a}{\partial p_E}\\[3mm]
K^c & = & p^Tg +p_Sg_S + p_Eg_E, \quad & g = \frac{\partial K^c}{\partial p}, \, g_{S} = \frac{\partial K^c}{\partial p_S}, \, g_E = \frac{\partial K^c}{\partial p_E},
\end{array}
\eq
where the functions $f,f_S,f_E$, as well as the elements of the $m$-dimensional row vectors of functions $g,g_S,g_E$, are all homogeneous of degree $0$. Now recall the energy representation \eqref{enerep} of the Lagrangian submanifold $\cL$ describing the state properties of the system
\bq
\label{state1}
\cL=\{(q,S,E,p,p_S,p_E) \mid E= E(q,S), p=-p_E \frac{\partial E}{\partial q}(q,S),  p_S=-p_E \frac{\partial E}{\partial S}(q,S) \}
\eq
By substitution of \eqref{state1} in \eqref{KaKc} it follows that $K$ restricted to $\cL$ is zero for all $u$ if and only if
\bq
\begin{array}{rcl}
\left(-p_E \frac{\partial E}{\partial q}f -p_E \frac{\partial E}{\partial S}f_S + p_Ef_E \right) |_{\cL} & = & 0 \\[2mm]
\left(-p_E \frac{\partial E}{\partial q}g -p_E \frac{\partial E}{\partial S}g_S + p_Eg_E \right) |_{\cL} & =& 0
\end{array}
\eq
for all $p_E$, or equivalently
\bq
\left(\frac{\partial E}{\partial q}f + \frac{\partial E}{\partial S}f_S\right) |_{\cL} = f_E |_{\cL}, \quad
\left(\frac{\partial E}{\partial q}g +\frac{\partial E}{\partial S}g\right) |_{\cL} = g_E  |_{\cL} 
%\end{array}
\eq
This leads to the following {\it additional} requirements on the homogeneous function $K^a$.
The {\it First Law of thermodynamics} (total energy preservation) requires that the uncontrolled ($u=0$) dynamics preserves energy, implying that
\bq
f_E |_{\cL}=0
\eq
Furthermore, the {\it Second Law of thermodynamics} ('increase of entropy') leads to the following requirement. Writing out $K |_{\cL}=0$ in the {\it entropy representation} \eqref{entrep} of $\cL$ amounts to 
\bq
% \begin{array}{l}
\left(\frac{\partial S}{\partial q}f + \frac{\partial S}{\partial E}f_E\right) |_{\cL} = f_S |_{\cL}, \quad 
\left(\frac{\partial S}{\partial q}g +\frac{\partial S}{\partial E}g_E\right) |_{\cL} = g_S  |_{\cL} 
%\end{array}
\eq
Plugging in the earlier found requirement $f_E |_{\cL}=0$ this reduces to
\bq
\frac{\partial S}{\partial q}f |_{\cL} = f_S |_{\cL}, \quad
\left(\frac{\partial S}{\partial q}g +\frac{\partial S}{\partial E}g_E\right) |_{\cL} = g_S  |_{\cL} 
\eq
Finally, since for $u=0$ the entropy is {\it non-decreasing} this implies the following additional requirement
\bq
f_S |_{\cL}\geq0
\eq
%Furthermore, in the standard case of a port representing {\it power-flow} we may define a vector of outputs $y$ (conjugate to the input vector $u$) as $y:=g_E|_{\cL}$
%
All this leads to the following geometric formulation of a {\it port-thermodynamic system}.
\begin{Definition}[Port-thermodynamic system]\label{portthermo}
Consider the space of extensive variables $Q^e=Q \times \mR \times \mR$, and the thermodynamic phase space $\mP(T^*Q^e)$. A port-thermodynamic system on $\mP(T^*Q^e)$ is defined as a pair $(\cL,K)$, where the homogeneous Lagrangian submanifold $\cL \subset \T^*Q^e$ specifies the state properties. The dynamics is given by the homogeneous Hamiltonian dynamics with parametrized homogeneous Hamiltonian $K:= K^a + K^cu : \T^*Q^e \to \mR, \, u \in \mR^m,$ in the form \eqref{KaKc}, with $K^a,K^c$ zero on $\cL$, and the internal Hamiltonian $K^a$ satisfying (corresponding to the First and Second Law of thermodynamics)
\bq
\label{firstsecond}
f_E |_{\cL}=0, \quad f_S |_{\cL}\geq0
\eq
This means that in energy representation \eqref{enerep}
\bq
\left(\frac{\partial E}{\partial q}f + \frac{\partial E}{\partial S}f_S\right) |_{\cL} = 0, \quad
\left(\frac{\partial E}{\partial q}g +\frac{\partial E}{\partial S}g_S\right) |_{\cL} = g_E  |_{\cL} 
\eq
and, in entropy representation \eqref{entrep}
\bq
\frac{\partial S}{\partial q}f |_{\cL} = f_S |_{\cL} \geq 0, \quad
\left(\frac{\partial S}{\partial q}g +\frac{\partial S}{\partial E}g_E\right) |_{\cL} = g_S  |_{\cL}
\eq
Furthermore, the {\it power-conjugate outputs} $y_p$ of the port-thermodynamic system $(\cL,K)$ are defined as the row-vector
\bq
y_p :=g_E|_{\cL}
\eq
\end{Definition}
Since by Euler's theorem (Theorem \ref{Euler}) all expressions $f,f_S,f_E,g,g_S,g_E$ are homogeneous of degree $0$, they project to functions on the thermodynamic phase space $\mP(T^*Q^e)$. Hence the dynamics and the output equations are equally well-defined on the Legendre submanifold $L \subset \mP(T^*Q^e)$.
\begin{Remark}
Definition \ref{portthermo} can be generalized to the compartmental situation $Q^e=Q \times \mR^{m_S} \times \mR^{m_E}$, by modifying \eqref{firstsecond} into the condition 
\bq
\label{firstsecond1}
\sum_{i=1}^{m_E} f_{E_i} |_{\cL}=0, \quad \sum_{j=1}^{m_S} f_{S_j} |_{\cL}\geq0 ,
\eq
corresponding to {\it total} energy conservation, and {\it total} entropy increase.
\end{Remark}
Note that as a consequence of the above definition of a port-thermodynamic system
\bq
\frac{d}{dt}E|_{\cL}= y_pu,
\eq
expressing that the increase of total energy of the thermodynamic system is equal to the energy supplied to the system by the environment.

\medskip

Defining the vector of outputs as being {\it power-conjugate} to the input vector $u$ is the most common option for defining an interaction port (in this case properly called a {\it power-port}) of the thermodynamic system. Nevertheless, there are other possibilities as well.
Indeed, a port representing {\it rate of entropy flow} is obtained by defining the alternative output $y_{re}$ as
\bq
y_{re}:=g_S|_{\cL},
\eq
which is {\it entropy-conjugate} to the input vector $u$,  
This leads instead to the {\it rate of entropy} balance
\bq
\frac{d}{dt}S|_{\cL}= y_{re}u +f_S |_{\cL},
\eq
where the second, non-negative, term on the right-hand side is the {\it internal rate of entropy production}.

\begin{Remark}
From the point of view of {\it dissipativity theory} \cite{willems, Sch17} this means that any port-thermodynamical system, with inputs $u$ and outputs $y_p, y_{re}$, is cyclo-lossless with respect to the supply rate $y_pu$, and cyclo-passive with respect to the supply rate $y_{re}u$.
\end{Remark}
\begin{Remark}
An extension to Definition \ref{portthermo} is to consider a non-affine dependence of $K$ on $u$, i.e., a general function $K: \T^*Q^e \times \mR^m \to \mR$ that is homogeneous in $p^e$. See also the formulation of {\it Hamiltonian input-output systems}, starting with \cite{brockett} and continued in e.g. \cite{vanderschaftMST, schaftcrouch, vds89}. 
\end{Remark}
\begin{Remark}
In case $f,f_S,f_E,g,g_S,g_E$ do not depend on $p^e$ (and therefore are trivially homogeneous of degree $0$), they actually define {\it vector fields} on the space of extensive variables $Q^e$ (since they transform under a coordinate change for $Q^e$ as vector fields). In this case the dynamics on $\cL$ is equal to the Hamiltonian lift of the dynamics on $Q^e$ to $T^*Q^e$ (with respect to the standard projection $p: T^*Q^e \to Q^e$), just like in \cite{schaftcrouch}.
\end{Remark}
It is of interest to note that, as illustrated by the following examples, the Hamiltonian $K$ generating the dynamics on $\cL$ is {\it dimensionless}; i.e., its values do not have a physical dimension. Physical dimensions {\it do} appear by dividing the homogeneous expression by one of the co-extensive variables. 

%\begin{remark}
%The geometric formulation of mechanical systems has spurred {\it symplectic geometry}; see e.g. the classical textbooks \cite{arnold}, \cite{abraham}, \cite{libermann}. Symplectic geometry was also underlying the formulation of {\it Hamiltonian input-output systems}, starting with the ground-breaking paper \cite{brockett} and continued in e.g. \cite{vanderschaftMST, schaftcrouch, vds89}. By generalizing symplectic and Poisson structures to Dirac structures, and by emphasizing port-based modeling of multi-physics systems, this also led to the theory of {\it port-Hamiltonian systems}.
%\end{remark}

\subsection{Examples of port-thermodynamic systems}\label{subsec:ex}
\begin{Example}[Heat compartment]\label{heatcomp}
Consider a simple thermodynamic system in a compartment, allowing for heat exchange with its environment. Its thermodynamic properties are described by the extensive variables $S$ (entropy) and $E$ (internal energy), with $E$ expressed as a function $E=E(S)$ of $S$. Its state properties (in energy representation) are given by the homogeneous Lagrangian submanifold
\bq
\cL= \{(S,E,p_S,p_E) \mid E=E(S), p_S=-p_EE'(S) \},
\eq
corresponding to the generating function $-p_EE(S)$. Since there is no internal dynamics $K^a$ is absent. Hence, taking $u$ as the {\it rate of entropy flow} corresponds to the homogeneous Hamiltonian $K=K^cu$ with
\bq
K^c=p_S + p_E E'(S), 
\eq
which is zero on $\cL$. This yields on $\cL$ the dynamics (entailing both the entropy and energy balance)
\bq
\begin{array}{rcllrcl}
\dot{S} &=& u  \quad & \dot{p}_S &=& -p_EE''(S)u \\[2mm]
\dot{E} &=&E'(S)u \quad & \dot{p}_E &=& 0,
\end{array}
\eq
with power-conjugate output $y_p$ equal to the temperature $T=E'(S)$. 
Defining the homogeneous coordinate $\gamma=-\frac{p_S}{p_E}$ leads to the contact Hamiltonian $\widehat{K}^c= E'(S) - \gamma$ on $\mP(T^*\mR^2)$, and the Legendre submanifold $L $
\bq
L= \{(S,E,\gamma) \in \mP(T^*\mR^2) \mid E=E(S), \gamma=E'(S) \}
\eq
The resulting contact dynamics on $L$ is equal to the projected dynamics $\pi_*X_K=X_{\widehat{K}}$ given as
\bq
\begin{array}{rcl}
\dot{S} &=& u\\[2mm]
\dot{E} &=&E'(S)u\\[2mm]
\dot{\gamma} & = & - \frac{\dot{p}_S}{p_E} = E''(S)u
\end{array}
\eq
Here, the third equation corresponds to the energy balance in terms of the temperature dynamics. Note that $E''(S)= \frac{T}{C}$, with $C$ the heat capacitance of the fixed volume. 

Alternatively, if instead we take the incoming {\it heat flow} as input $v$, then the Hamiltonian is given by
\bq
K=(p_S \frac{1}{E'(S)} + p_E)v,
\eq
leading to the 'trivial' power-conjugate output $y_p=1$, and to the rate of entropy conjugate output $y_{re}$ given by the reciprocal temperature $y_{re}=\frac{1}{T}$.
\end{Example}

\begin{Example}[Heat exchanger]\label{heatexch}
Consider two heat compartments as in Example \ref{heatcomp}, exchanging a heat flow through an interface according to Fourier's law. The extensive variables are $S_1,S_2$ (entropies of the two compartments) and $E$ (total internal energy). The state properties are described by the homogeneous Lagrangian submanifold
\bq
%\begin{array}{rcl}
\cL  = \{(S_1,S_2,E,p_{S_1},p_{S_2},p_E) \mid E=E_1(S_1) +E_2(S_2), 
%\\[2mm]
p_{S_1}=-p_EE_1'(S_1), p_{S_2}=-p_EE_2'(S_2) \},
%\end{array}
\eq
corresponding to the generating function $-p_E\left(E_1(S_1) + E_2(S_2)\right)$, with $E_1,E_2$ the internal energies of the two compartments. Denoting the temperatures $T_1=E_1'(S_1),$ $ T_2=E_2'(S_2)$, the internal dynamics of the two-component thermodynamic system corresponding to Fourier's law is given by the Hamiltonian
\bq
\label{heatexchK}
K^a = \lambda (\frac{1}{T_1} - \frac{1}{T_2})(p_{S_1}T_2 - p_{S_2}T_1),
\eq
with $\lambda$ Fourier's conduction coefficient. Note that the total entropy on $\cL$ satisfies
\bq
\dot{S}_1 + \dot{S}_2 = \lambda (\frac{1}{T_1} - \frac{1}{T_2})(T_2 - T_1) \geq 0,
\eq
in accordance with \eqref{firstsecond1}.
We will revisit this example in the context of {\it interconnection} of thermodynamic systems in Examples \ref{ex:heatexch} and \ref{ex:heatexch1}.
\end{Example}

\begin{Example}[Mass-spring-damper system]\label{msd}
Consider a mass-spring-damper system in one-dimensional motion, composed of a mass $m$ with momentum $\pi$, linear spring with stiffness $k$ and extension $z$, and linear damper with viscous friction coefficient $d$. In order to take into account the thermal energy and the entropy production arising from the heat produced by the damper, the variables of the mechanical system are augmented with an entropy variable $S$ and internal energy $U(S)$. (For instance, if the system is isothermal, i.e. in thermodynamic equilibrium with a thermostat at temperature $T_0$, the internal energy is $U(S) = T_0S$.) This leads to the total set of extensive variables $z$, $\pi$, $S$, $E=\frac{1}{2}kz^2 + \frac{\pi^2}{2m} + U(S)$ (total energy). The state properties of the system are described by the Lagrangian submanifold $\cL$ with generating function (in energy representation)
\bq
-p_E\left(\frac{1}{2}kz^2 + \frac{\pi^2}{2m} + U(S)\right)
\eq
This defines the state properties
\bq
%\begin{array}{rcl}
\cL = \{(z,\pi,S,E,p_z,p_{\pi},p_S,p_E) | E=\frac{1}{2}kz^2 + \frac{\pi^2}{2m} + U(S), 
%\\[2mm]
p_z=-p_Ekz, p_{\pi}= -p_E \frac{\pi}{m},p_S=-p_EU'(S) \}
%\end{array}
\eq
The dynamics is given by the homogeneous Hamiltonian
\bq
\label{msdK}
K= p_z\frac{\pi}{m} +p_{\pi}\left(-kz -d\frac{\pi}{m}\right) +p_S \frac{d (\frac{\pi}{m})^2}{U'(S)} + \left(p_{\pi} + p_E \frac{\pi}{m}\right)u,
\eq
where $u$ is an external force. The power-conjugate output $y_p=\frac{\pi}{m}$ is the velocity of the mass.
\end{Example}

\begin{Example}[Gas-piston-damper system]
Consider a gas in an adiabatically isolated cylinder closed by a piston. Assume that the thermodynamic properties of the system are covered by the
properties of the gas. (For an extended model see \cite{favache10}, Section 4.)
Then the system is analogous to the previous example, replacing $z$ by volume $V$ and the partial energy $\frac{1}{2}kz^2 + U(S)$ by an expression $U(V,S)$ for the internal energy of the gas. The dynamics of a force-actuated gas-piston-damper system is defined by the Hamiltonian
\bq
K= p_z\frac{\pi}{m} +p_{\pi}\left(-\frac{\partial U}{\partial V} -d\frac{\pi}{m}\right) +p_S \frac{d (\frac{\pi}{m})^2}{\frac{\partial U}{\partial S}} + \left(p_{\pi} + p_E \frac{\pi}{m}\right)u,
\eq
where the power-conjugate output $y_p=\frac{\pi}{m}$ is the velocity of the piston.
\end{Example}

\begin{Example}[Port-Hamiltonian systems as port-thermodynamic systems]
Example \ref{msd} can be extended to any input-state-output port-Hamiltonian system \cite{maschkevds, vanderschaftmaschkearchive, vds14}
\bq
\label{pH}
\begin{array}{rcl}
\dot{x} & = & J(x)e - R(e) +G(x)u , \quad e=\frac{\partial H}{\partial x}(x), \; J(x)=-J^T(x) \\[2mm]
y& = &G^T(x)e
\end{array}
\eq
on a state space manifold $x \in \mathcal{X}$, with inputs $u \in \mR^m$, outputs $y \in \mR^m$, Hamiltonian $H$ (equal to the stored energy of the system), and dissipation $R(e)$ satisfying $e^TR(e)\geq0$ for all $e$.
Including {\it entropy} $S$ as an {\it extra} variable, along with an {\it internal energy} $U(S)$ (for example in the isothermal case $U(S)=T_0S$), the state properties of the port-Hamiltonian system are given by the homogeneous Lagrangian submanifold $\cL \subset T^*(\mathcal{X} \times \mR^2)$ defined as
\bq
%\begin{array}{rcl}
\cL =  \{(x,S,E,p,p_S,p_E) \mid E(x,S)=H(x) + U(S), 
%\\[2mm]
\, p=-p_E\frac{\partial H}{\partial x}(x),p_S=-p_EU'(S) \},
%\end{array}
\eq
with generating function $-p_E\left(H(x) + U(S)\right)$. The Hamiltonian $K$ is given by (using the shorthand notation $e=\frac{\partial H}{\partial x}(x)$)
\bq
K(x,S,E,p,p_S,p_E)=p^T \left(J(x)e - S(e) +G(x)u\right) - p_S\frac{e^TR(e)}{U'(S)} +p_Ee^TG(x)u
\eq
reproducing on $\cL$ the dynamics \eqref{pH} with outputs $y_p=y$. 
Note that in this thermodynamic formulation of the port-Hamiltonian system the energy-dissipation term $e^TR(e)$ in the power-balance $\frac{d}{dt}H=-e^TR(e) +y^Tu$ is compensated by equal increase of the internal energy $U(S)$, thus leading to conservation of the total energy $E(x,S)=H(x) + U(S)$.
\end{Example}

\subsection{Controllability of port-thermodynamic systems}\label{subsec:cont}
In this subsection we will briefly indicate how the {\it controllability} properties of the port-thermodynamic system $(\cL,K)$ can be directly studied in terms of the homogeneous Hamiltonians $K^a$ and $K^c_j,\, j=1, \cdots,m,$ and their {\it Poisson brackets}. First we note that by Proposition \ref{app:poisson} the Poisson brackets of these homogeneous Hamiltonians are again homogeneous. Secondly, we recall the well-known correspondence \cite{libermann,arnold, abraham} between Poisson brackets of Hamiltonians $h_1,h_2$ and Lie brackets of the corresponding Hamiltonian vector fields
\bq
[X_{h_1},X_{h_2}] =X_{\{h_1,h_2\}}
\eq
In particular, this property implies that if the homogeneous Hamiltonians $h_1,h_2$ are zero on the homogeneous Lagrangian submanifold $\cL$, and thus by Proposition \ref{propinv} the homogeneous Hamiltonian vector fields $X_{h_1},X_{h_2}$ are tangent to $\cL$, then also $[X_{h_1},X_{h_2}]$ is tangent to $\cL$, and therefore the Poisson bracket $\{h_1,h_2\}$ is also zero on $\cL$. Furthermore, with respect to the projection to the corresponding Legendre submanifold $L$, we note the following property of homogeneous Hamiltonians
\bq
\widehat{\{h_1,h_2\}}= \{\widehat{h}_1,\widehat{h}_2\},
\eq
where the bracket on the right-hand side is the {\it Jacobi bracket} \cite{libermann,arnold} of functions on the contact manifold $\mP(T^*Q^e)$.
This leads to the following analysis of the {\it accessibility algebra} \cite{nvds} of a port-thermodynamic system, characterizing its controllability .
\begin{Proposition}
Consider a port-thermodynamic system $(\cL,K)$ on $\mP(T^*Q^e)$ with homogeneous $K:= K^a + \sum_{j=1}^mK_j^cu_j : \T^*Q^e \to \mR$, zero on $\cL$. Consider the algebra $\cP$ (with respect to the Poisson bracket) generated by $K^a, K_j^c, j=1, \cdots,m$, consisting of homogeneous functions that are zero on $\cL$, and the corresponding algebra $\widehat{\cP}$ generated by $\widehat{K}^a, \widehat{K}_j^c, j=1, \cdots,m,$ on $L$. The accessibility algebra \cite{nvds} is spanned by all contact vector fields $X_{\widehat{h}}$ on $L$, with $\widehat{h}$ in the algebra $\widehat{\cP}$.
It follows that the port-thermodynamic system $(\cL,K)$ is locally accessible \cite{nvds} if the dimension of the co-distribution $d \widehat{\cP}$ on $L$ defined by the differentials of $\widehat{h}$, with $h$ in the Poisson algebra $\cP$, is equal to the dimension of $L$. Conversely, if the system is locally accessible then the co-distribution $d \widehat{\cP}$ on $L$ has dimension equal to the dimension of $L$ almost everywhere on $L$.
\end{Proposition}
Similar statements can be made with respect to local strong accessibility of the port-thermodynamic system; see the theory exposed in \cite{nvds}.

\section{Interconnections of port-thermodynamic systems}\label{sec:int}
 In this section we study the geometric formulation of {\it interconnection} of port-thermodynamic systems through their ports, in the spirit of the compositional theory of port-Hamiltonian systems \cite{maschkevds, vanderschaftmaschkearchive, vds14, Sch17}. We will concentrate on the case of {\it power-port} interconnections of port-thermodynamic systems, corresponding to power flow exchange (with total power conserved). This is the standard situation in physical network modeling of interconnected systems, in particular in port-based modeling theory, see e.g. \cite{golo}. At the end of this section we will make some remarks about other types of interconnection; in particular interconnection by exchange of {\it rate of entropy}. 
 
Consider two port-thermodynamic systems with extensive and co-extensive variables
\bq
(q_i,p_i,S_i,p_{S_i},E_i,p_{E_i}) \in T^*Q^e_i=T^*Q_i \times T^*\mR_i \times T^*\mR_i, \quad i=1,2, 
\eq
and Liouville one-forms $\alpha_i= p_idq_i + p_{S_i}dS_i  + p_{E_i}dE_i, \, i=1,2$. With the homogeneity assumption in mind, impose the following constraint on the co-extensive variables
\bq
p_{E_1} = p_{E_2}=: p_E
\eq
This leads to the summation of the one-forms $\alpha_1$ and $\alpha_2$ given by
\bq
\alpha_{\mathrm{sum}}:= p_1dq_1 + p_2dq_2 + p_{S_1}dS_1 + p_{S_2}dS_2 + p_Ed(E_1 + E_2)
\eq
on the {\it composed space} defined as
\bq
%\begin{array}{rcl}
T^*Q^e_1 \circ T^*Q^e_2 :=  \{(q_1,p_1, q_2,p_2, S_1,p_{S_1}, S_2,p_{S_2},E,p_{E}) 
%\\[2mm]
\in T^*Q_1\times T^*Q_2  \times T^*\mR \times T^*\mR \times T^*\mR \}
%\end{array}
\eq
Leaving out the zero-section $p_1=0,p_2=0, p_{S_1}=0, p_{S_2}=0,p_{E}=0$, this space will be denoted by $\T^*Q^e_1 \circ \T^*Q^e_2$, and will serve as the space of extensive and co-extensive variables for the interconnected system. Furthermore, it defines the projectivization $\mP(T^*Q^e_1 \circ T^*Q^e_2)$, which serves as the composition (through $E_i,p_{E_i}, i=1,2$) of the two projectivizations $\mP(T^*Q^e_i), i=1,2$.

Let the state properties of the two systems be defined by homogeneous Lagrangian submanifolds
\bq
\cL_i \subset T^*Q_i \times T^*\mR_i \times T^*\mR_i, \quad i=1,2,
\eq
with generating functions $-p_{E_i}E_i(q_i,S_i), i=1,2$. Then the state properties of the composed system are defined by the composition
\bq
\label{circ}
%\begin{array}{rcl}
\cL_1 \circ \cL_2 :=  \{(q_1,q_2,p_1,p_2, S_1,p_{S_1}, S_2,p_{S_2},E,p_{E} \mid E=E_1 + E_2, 
\\[2mm]
\, (q_i,p_i,S_i,p_{S_i},E_i,p_{E_i}) \in \cL_i, \; i=1,2 \},
%\end{array}
\eq
with generating function $-p_{E} \left(E_1(q_1,S_1) + E_2(q_2,S_2)\right)$.

Furthermore, consider the dynamics on $\cL_i$ defined by the Hamiltonians $K_i =K_i^a + K_i^cu_i, i=1,2$. {\it Assume} that $K_i$ does {\it not} depend on the energy variable $E_i, i=1,2$. Then the sum $K_1 + K_2$ is well-defined on $\cL_1 \circ \cL_2$ for all $u_1,u_2$. This defines a {\it composite port-thermodynamic system}, with entropy variables $S_1,S_2$, total energy variable $E$, inputs $u_1,u_2$ and state properties defined by $\cL_1 \circ \cL_2$. 

Next, consider the {\it power-conjugate} outputs $y_{p1},y_{p2}$; in the sequel simply denoted by $y_1,y_2$. By imposing {\it interconnection constraints} on the power-port variables $u_1,u_2,y_1,y_2$ satisfying the {\it power-preservation} property
\bq
\label{intercon}
y_1u_1 + y_2u_2=0,
\eq
then yields an interconnected dynamics on $\cL_1 \circ \cL_2$, which is energy-conserving (the $p_E$-term in the expression for $K_1 + K_2$ is zero by \eqref{intercon}.). This is summarized in the following proposition.

\begin{Proposition}
\label{prop:inter}
Consider two port-thermodynamic systems $(\cL_i,K_i)$ with spaces of extensive variables $Q^e_i$, $i=1,2$. Assume that $K_i$ does not depend on $E_i$, $i=1,2$. Then $(\cL_1 \circ \cL_2,K_1 + K_2)$, with $\cL_1 \circ \cL_2$ given in \eqref{circ}, defines a composite port-thermodynamic system with inputs $u_1,u_2$ and outputs $y_1,y_2$. By imposing interconnection constraints on $u_1,u_2, y_1,y_2$ satisfying \eqref{intercon} an autonomous (no inputs) port-thermodynamic system is obtained.
\end{Proposition}
\begin{Remark}
The interconnection procedure can be extended to the case of an additional {\it open} power-port with input vector $u$ and output row vector $y$, by replacing \eqref{intercon} by power-preserving interconnection constraints on $u_1,u_2,u,y_1,y_2,y,$ satisfying
\bq
y_1u_1 + y_2u_2 +yu=0
\eq
\end{Remark}
Proposition \ref{prop:inter} is illustrated by the following examples.
\begin{Example}[Mass-spring-damper system]
We will show how the thermodynamic formulation of the system as detailed in Example \ref{msd} also results from the interconnection of the three subsystems: mass, spring and damper.\\
I. {\it Mass subsystem} (leaving out irrelevant entropy). The state properties are given by
\bq
\cL_m= \{(\pi,\kappa,p_{\pi},p_{\kappa}) \mid \kappa=\frac{\pi^2}{2m}, p_{\pi} = -p_{\kappa} \frac{\pi}{m} \},
\eq
with energy $\kappa$ (kinetic energy), and dynamics generated by the Hamiltonian
\bq
K_m= (p_{\kappa} \frac{\pi}{m} +p_{\pi})u_m,
\eq
corresponding to $\dot{\pi}=u_m, y_m=\frac{\pi}{m}$.\\
II. {\it Spring subsystem} (again leaving out irrelevant entropy). The state properties are given by
\bq
\cL_s= \{(z,P,p_z,p_P) \mid P=\frac{1}{2},kq^2 p_z = -p_P kz \},
\eq
with energy $P$ (spring potential energy), and dynamics generated by the Hamiltonian
\bq
K_s= (p_P kz +p_{z})u_s,
\eq
corresponding to $\dot{z}=u_s, y_s=kz$.\\
III. {\it Damper subsystem}.
The state properties are given by
\bq
\cL_d=\{ (S,U) \mid U=U(S), p_S=-p_UU'(S) \},
\eq
involving the entropy $S$ and an internal energy $U(S)$.
The dynamics of the damper subsystem is generated by the Hamiltonian
\bq
K_d= (p_U + p_{S} \frac{1}{U'(S)})du_d^2
\eq
with $d$ the damping constant, and power-conjugate output 
\bq
y_d:= du_d
\eq
equal to the damping force.

Finally interconnect, in a power-preserving way, the three subsystems to each other via their power-ports $(u_m,y_m), (u_s,y_s), (u_d,y_d)$ as
\bq
u_m= -y_s - y_d, \; u_s=y_m=u_d
\eq
This results (after setting $p_{\kappa}=p_P=p_U=:p$) in the interconnected port-thermodynamic system with total Hamiltonian $K_m + K_s + K_d$ given as
\bq
\begin{array}{l}
(p \frac{\pi}{m} +p_{\pi})u_m + (p kz +p_{z})u_s + (p + p_{S} \frac{1}{U'(S)})du_d^2=\\[2mm]
(p \frac{\pi}{m} +p_{\pi})(-kz -d\frac{\pi}{m}) + (p kz +p_{z})\frac{\pi}{m} + (p + p_{S} \frac{1}{U'(S)})d(\frac{\pi}{m})^2 =\\[2mm]
p_z\frac{\pi}{m} +p_{\pi}(-kz -d\frac{\pi}{m}) +p_S \frac{d (\frac{\pi}{m})^2}{U'(S)},
\end{array}
\eq
which is equal to the Hamiltonian for $u=0$ as obtained before in Example \ref{msd}, eqn. \eqref{msdK}.
\end{Example}

\begin{Example}[Heat exchanger]\label{ex:heatexch}
Consider two heat compartments as in Example \ref{heatcomp}, with state properties
\bq
\cL_i= \{(S_i,E_i,p_{S_i},p_{E_i}) \mid E_i=E_i(S_i), p_{S_i}=-p_{E_i}E_i'(S) \},\; i=1,2.
\eq
The dynamics is given by the Hamiltonians
\bq
K_i=(p_{E_i} + p_{S_i}\frac{1}{T_i})v_i, \quad \; T_i=E_i'(S_i), \quad i=1,2,
\eq
with $v_1,v_2$ the incoming heat flows, and power-conjugate outputs $y_1, y_2$, which both are equal to $1$. Consider the power-conserving interconnection
\bq
\label{h1}
v_1=-v_2= \lambda (T_2 - T_1),
\eq
with $\lambda$ the Fourier heat conduction coefficient.
Then the Hamiltonian of the interconnected port-thermodynamical system is given by
\bq
K_1 + K_2= \lambda (T_2 - T_1)(\frac{p_{S_1}}{T_1} - \frac{p_{S_2}}{T_2}) ,
\eq
which equals the Hamiltonian \eqref{heatexchK} as obtained in Example \ref{heatexch}.
\end{Example}

Apart from power-port interconnections as above, we may also define other types of interconnection, {\it not} corresponding to exchange of rate of energy (power), but instead to exchange of rate of other extensive variables. In particular, an interesting option is to consider interconnection via rate of {\it entropy} exchange. This can be done in a similar way, by considering, instead of the variables $E_i,p_{E_i}, i=1,2,$ as above, the variables $S_i,p_{S_i}, i=1,2$. Imposing alternatively the constraint $p_{S_1} = p_{S_2}=: p_S$, yields a similar composed space of extensive and co-extensive variables, as well as a similar composition $\cL_1 \circ \cL_2$ of the state properties. By assuming in this case that the Hamiltonians $K_i$ do not depend on the entropies $S_i, i=1,2$, and by imposing interconnection constraints on $u_1,u_2$ and the 'rate of entropy' conjugate outputs $y_{re1}, y_{re2}$, leads again to an interconnected port-thermodynamic system. Note however that while it is natural to assume conservation of total energy for interconnection of two systems via their power-ports, in the alternative case of interconnecting through rate of entropy ports the total entropy may not be conserved, but actually increasing.

\begin{Example}
\label{ex:heatexch1}
As an alternative to the previous Example \ref{ex:heatexch}, where the heat exchanger was modelled as the interconnection of two heat compartments via power-ports, consider the same situation but now with outputs $y_i$ being 'rate of entropy conjugate' to $v_i$, i.e., equal (cf. the end of Example \ref{heatcomp}) to the reciprocal temperatures $\frac{1}{T_i}$ with $T_i=E'(S_i)$, $i=1,2$. This results in interconnecting the two heat compartments as, equivalently to \eqref{h1},
\bq
v_1=-v_2= \lambda (\frac{1}{y_2} - \frac{1}{y_1})
\eq
This interconnection is not total entropy conserving, but instead satisfies $y_1v_1 +y_2v_2= \lambda (\frac{1}{y_2} - \frac{1}{y_1})(y_1-y_2) \geq0$, corresponding to increase of total entropy.
\end{Example}

%%%%%%%%%%%%%%%%%%%%%%%%%%%%%%%%%%%%%%%%%%
\section{Discussion}
While the state properties of thermodynamic systems were geometrically formulated since the 1970s through the use contact geometry, in particular by Legendre submanifolds, the geometric formulation of non-equilibrium thermodynamic processes has remained more elusive. Taking up the symplectization point of view on thermodynamics as successfully initiated in \cite{balian}, the present paper developed a geometric framework based on the description of non-equilibrium thermodynamic processes by Hamiltonian dynamics on the symplectized thermodynamic phase space generated by Hamiltonians that are homogeneous of degree one in the co-extensive variables, and culminating in the definition of {\it port-thermodynamic systems} in Section 4.1. Furthermore, it was shown in Section 3 how the symplectization point of view provides an intrinsic definition of a metric that is overarching the locally defined metrics of Weinhold and Ruppeiner. The correspondence between objects in contact geometry and corresponding homogeneous objects in symplectic geometry turned to be very effective. An additional benefit of symplectization is the simplicity of the expressions and computations in the standard Hamiltonian context, as compared to those in contact geometry, which was exemplified in the initial controllability study in Section 4.3.

Although the presented simple examples demonstrated the potential of this geometric formulation of thermodynamics for analysis and control, there are many questions to be posed, and many application contexts to be addressed. From an analysis point of view it is to be seen if the identification of the Hamiltonian structure will be as important as in the analysis of mechanical systems. From a control point of view, while initial steps were provided for the study of controllability, the theory developed so far suggests many open problems, including the stabilization of thermodynamic processes and its applications.

%Authors should discuss the results and how they can be interpreted in perspective of previous studies and of the working hypotheses. The findings and their implications should be discussed in the broadest context possible. Future research directions may also be highlighted.
%
%%%%%%%%%%%%%%%%%%%%%%%%%%%%%%%%%%%%%%%%%%
%\section{Conclusions}
%
%This section is not mandatory, but can be added to the manuscript if the discussion is unusually long or complex.
%
%%%%%%%%%%%%%%%%%%%%%%%%%%%%%%%%%%%%%%%%%%
\vspace{6pt}

\appendixtitles{yes} %Leave argument "no" if all appendix headings stay EMPTY (then no dot is printed after "Appendix A"). If the appendix sections contain a heading then change the argument to "yes".
\appendixsections{one} %Leave argument "multiple" if there are multiple sections. Then a counter is printed ("Appendix A"). If there is only one appendix section then change the argument to "one" and no counter is printed ("Appendix").

\appendix
\section{Homogeneity of functions, of Hamiltonian vector fields, and of Lagrangian submanifolds}
In this section we throughout use, for notational simplicity, the notation $M$ instead of $Q^e$. Furthermore, we let $\dim M = n+1$, denote coordinates for $M$ by $q=( q_0,q_1, \cdots,q_n)$, and co-tangent bundle coordinates for $T^*M$ by $(q,p)=(q_0,q_1, \cdots,q_n, p_0, p_1, \cdots, p_n)$.

Fundamental will be the notion of {\it homogeneity} in the variables $p$.
\begin{Definition}
Let $r \in \mZ$. A function $K: \T^*M \to \mR$ is called homogeneous of degree $r$ (in the variables $p=(p_0, p_1 \cdots, p_n)$) if
\bq
%\begin{array}{l}
K(q_0,q_1, \cdots,q_n, \lambda p_0, \lambda p_1, \cdots, \lambda p_n) = 
%\\[2mm]
\lambda^r K(q_0,q_1, \cdots,q_n, p_0, p_1, \cdots, p_n), \quad \forall \lambda \neq 0
%\end{array}
\eq
\end{Definition}
Note that this definition is independent of the choice of cotangent-bundle coordinates $(q,p)$ for $\T^*M$.
\begin{Theorem}[Euler's homogeneous function theorem]\label{Euler}
A differentiable function $K: \T^*M \to \mR$ is homogeneous of degree $r$ (in the variables $p_0,p_1, \cdots, p_n$) if and only if
\bq
\sum_{i=0}^n p_i\frac{\partial K}{\partial p_i}(q,p)= rK(q,p), \quad \mbox{ for all } (q,p) \in \T^*M
\eq
Furthermore, if $K$ is homogeneous of degree $r$, then its derivatives $\frac{\partial K}{\partial p_i}, i=0,1,\cdots,n,$ are homogeneous of degree $r-1$.
\end{Theorem}
Geometrically, Euler's theorem can be equivalently formulated as follows. Recall that the {\it Hamiltonian vector field} $X_h$ on $T^*M$ with symplectic form $\omega=d \alpha$ corresponding to an arbitrary Hamiltonian $h: T^*M \to \mR$ is defined by $i_{X_h}\omega = -dh$. It is immediately verified that $h: T^*M \to \mR$ is homogeneous of degree $r$ iff 
\bq
\label{9}
\alpha (X_h) = r \, h
\eq
Furthermore, define the {\it Euler vector field} (also called {\it Liouville vector field}) $E$ on $T^*M$ as the unique vector field satisfying 
\bq
\label{9b}
d \alpha (E,\cdot) = \alpha
\eq
In co-tangent bundle coordinates $(q,p)$ for $T^*M$ the vector field $E$ is given as $\sum_{i=0}^n p_i \frac{\partial}{\partial p_i}$. One verifies that $h: T^*M \to \mR$ is homogeneous of degree $r$ iff 
\bq
\label{9a}
\mL_E h = r \, h
\eq
In the sequel we will only use homogeneity and Euler's theorem for $r=0$ and $r=1$. First, it is clear that physical variables defined on the contact manifold $\mP (T^*Q^e)$ correspond to functions on $\T^*Q^e$ which are {\it homogeneous of degree $0$}. On the other hand, as formulated in Proposition \ref{homvf}, a Hamiltonian vector field on $\T^*Q^e$ with respect to a Hamiltonian that is {\it homogeneous of degree $1$} projects to a contact vector field on the contact manifold $\mP (T^*Q)$. Such Hamiltonian vector fields are locally characterized as follows.
\begin{Proposition}\label{prop1}
If $h: T^*M \to \mR$ is homogeneous of degree $1$ then $X=X_h$ satisfies
\bq
\label{10}
\mL_{X}\alpha=0
\eq
Conversely, if the vector field $X$ satisfies \eqref{10} then $X=X_h$ for some locally defined Hamiltonian $h$ that is homogeneous of degree $1$.
\end{Proposition}
\begin{proof}
Note that by Cartan's formula $\mL_X=i_Xd + di_X$ (with $\mL$ denoting Lie-derivative, $d$ the exterior derivative, and $X$ any vector field). Hence
\bq
\label{11}
\mL_{X}\alpha = i_{X} d\alpha + d i_{X}\alpha =i_{X} d\alpha + d \left(\alpha (X)\right)
\eq
If $h$ is homogeneous of degree $1$ in $p$ then by \eqref{9} we have $\alpha (X_h) = h$, and thus $i_{X_h} d\alpha + d \alpha (X_h)= -dh +dh=0$, implying by \eqref{11} that $\mL_{X_h}\alpha=0$. Conversely, if $\mL_{X}\alpha=0$, then \eqref{11} yields $i_{X} d\alpha + d \left(\alpha (X)\right)=0$, implying that $X=X_h$, with $h=\alpha(X)$, which by \eqref{9} for $r=1$ is homogeneous of degree $1$.
\end{proof}
Summarizing, Hamiltonian vector fields with Hamiltonians that are homogeneous of degree $1$ are characterized by \eqref{10}; in contrast to general Hamiltonian vector fields $X$ on $T^*M$ which are characterized by the weaker property $\mL_{X} d\alpha=0$.

As before, in the sequel we will refer to Hamiltonians $h$ that are {\it homogeneous of degree} $1$ in the co-extensive variables $p$ simply as {\it homogeneous Hamiltonians}, and to Hamiltonian vector fields with a Hamiltonian that is homogeneous of degree $1$ in $p$ as {\it homogeneous Hamiltonian vector fields}. 

Similar statements as above can be made for homogeneous Lagrangian submanifolds (cf. Definition \ref{laghom}). Recall \cite{libermann, arnold,abraham} that a submanifold $\cL \subset T^*M$ is called a {\it Lagrangian submanifold} if the symplectic form $\omega:=d \alpha$ is zero on $\cL$, and $\dim \cL= \dim M$.
\begin{Proposition}\label{app:lag}
Consider the cotangent bundle $T^*M$ with its canonical one-form $\alpha$ and symplectic form $\omega:=d\alpha$. A submanifold $\cL \subset \T^*M$ is a {\it homogeneous Lagrangian submanifold} if and only if $\alpha$ restricted to $\cL$ is zero, and $\dim \cL= \dim M$.
\end{Proposition}
\begin{proof} First of all note the following. Recall the definition of the Euler vector field $E$ in \eqref{9b}. In co-tangent bundle coordinates $(q,p)$ for $T^*M$, the Euler vector field takes the form $E=\sum_{i=0}^n p_i \frac{\partial}{\partial p_i}$. Hence homogeneity of $\cL$ is equivalent to {\it tangency} of $E$ to $\cL$.\\
(If) By Palais' formula (see e.g. \cite{abraham}, Proposition 2.4.15)
\bq
d \alpha (X_0,X_1) = \mL_{X_0}(\alpha (X_1)) -  \mL_{X_1}(\alpha (X_0)) - \alpha \left( [X_0,X_1] \right)
\eq
for any two vector fields $X_0,X_1$. Hence, for any $X_1,X_2$ tangent to $\cL$ we obtain $d \alpha (X_0,X_1)=0$, implying that $d \alpha$ is zero restricted to $\cL$, and thus $\cL$ is a Lagrangian submanifold. Furthermore, by \eqref{9b}
\bq
\label{alpha}
d\alpha(E,X)=\alpha (X)=0,
\eq
for all vector fields $X$ tangent to $\cL$. Because $\cL$ is a Lagrangian submanifold this implies that $E$ is tangent to $\cL$ (since a Lagrangian submanifold is a {\it maximal} submanifold restricted to which $\omega=d \alpha$ is zero.) Hence $\cL$ is a homogeneous.\\
(Only if). If $\cL$ is homogeneous, then $E$ is tangent to $\cL$, and thus, since $\cL$ is Lagrangian, \eqref{alpha} holds for all vector fields $X$ tangent to $\cL$, implying that $\alpha$ is zero restricted to $\cL$.
\end{proof}

Regarding the {\it Poisson brackets} of Hamiltonian functions that are either homogeneous of degree $1$ or $0$ (in $p$), we have the following proposition.
\begin{Proposition}\label{app:poisson}
Consider the Poisson bracket $\{h_1,h_2\}$ of functions $h_1,h_2$ on $T^*M$ defined with respect to the symplectic form $\omega=d\alpha$. Then
\begin{enumerate}
\item[(a)]
If $h_1,h_2$ are both homogeneous of degree $1$, then also $\{h_1,h_2\}$ is homogeneous of degree $1$.
\item[(b)]
If $h_1$ is homogeneous of degree $1$, and $h_2$ is homogeneous of degree $0$, then $\{h_1,h_2\}$ is homogeneous of degree $0$.
\item[(c)]
If $h_1,h_2$ are both homogeneous of degree $0$, then $\{h_1,h_2\}$ is zero.
\end{enumerate}
\end{Proposition}
\begin{proof}
\begin{enumerate}
\item[(a)]
Since $h_1,h_2$ are both homogeneous of degree $1$ we have by Proposition \ref{prop1} $\mL_{X_{h_i}}\alpha=0, i=1,2$. Hence
\bq
%\begin{array}{c}
\mL_{X_{\{h_1,h_2\}}} \alpha = \mL_{[X_{h_1},X_{h_2}]} \alpha = 
\mL_{X_{h_1}}(\mL_{X_{h_2}}\alpha) - \mL_{X_{h_2}}(\mL_{X_{h_1}}\alpha) =0 ,
%\end{array}
\eq
implying by Proposition \ref{prop1} that $\{h_1,h_2\}$ is homogeneous of degree $1$.
\item[(b)]
$\alpha(X_{h_2})=0$, while by Proposition \ref{prop1} $\mL_{X_{h_1}}\alpha=0$, implying
\bq
0= \mL_{X_{h_1}} (\alpha(X_{h_2}))= 
(\mL_{X_{h_1}} \alpha)(X_{h_2}) + \alpha ([X_{h_1},X_{h_2}])= \alpha(X_{\{h_1,h_2\}}),
\eq
which means that $\{h_1,h_2\}$ is homogeneous of degree $0$.
\item[(c)]
First we note that for any $X_h$ with $h$ homogeneous of degree $0$, since $\alpha(X_{h})=0$,
\bq
\label{proofc}
\mL_{X_{h}}\alpha= i_{X_{h}} d \alpha + d (i_{X_{h}}\alpha)= -dh
\eq
Utilizing this property for $h_1$, we obtain, since $\alpha(X_{h_2})=0$,
\bq
\begin{array}{c}
0= \mL_{X_{h_1}} (\alpha(X_{h_2}))= (\mL_{X_{h_1}}\alpha)(X_{h_2}) + \alpha (X_{\{h_1,h_2\}}) = 
\\[2mm]
 - dh_1(X_{h_2}) + \alpha (X_{\{h_1,h_2\}}) = - \{h_1,h_2\} + \alpha (X_{\{h_1,h_2\}}),
\end{array}
 \eq
proving that $\{h_1,h_2\}$ is homogeneous of degree $1$. Hence by Proposition \ref{prop1} $\mL_{X_{\{h_1,h_2\}}} \alpha=0$, and thus
\bq
\begin{array}{c}
0= \mL_{X_{\{h_1,h_2\}}} \alpha = \mL_{[X_{h_1}, X_{h_2}]} \alpha = \mL_{X_{h_1}} \mL_{X_{h_2}}\alpha- \mL_{X_{h_2}}\mL_{X_{h_1}}\alpha=
\\[2mm]
\mL_{X_{h_1}}(-dh_2) - \mL_{X_{h_2}}(-dh_1) = -2 \{h_1,h_2\}
\end{array}
\eq
where in the fourth equality we use \eqref{proofc} for $h_1$ and $h_2$.
\end{enumerate}
\end{proof}

%\unskip
%\subsection{}
%The appendix is an optional section that can contain details and data supplemental to the main text. For example, explanations of experimental details that would disrupt the flow of the main text, but nonetheless remain crucial to understanding and reproducing the research shown; figures of replicates for experiments of which representative data is shown in the main text can be added here if brief, or as Supplementary data. Mathematical proofs of results not central to the paper can be added as an appendix.
%
%%\section{}
%All appendix sections must be cited in the main text. In the appendixes, Figures, Tables, etc. should be labeled starting with `A', e.g., Figure A1, Figure A2, etc. 


%=====================================
% References, variant A: internal bibliography
%=====================================
\reftitle{References}

%%%%%%%%%%%%%%%%%%%%%%%%%%%%%%%%%%%%%%%%%%
\end{document}